\documentclass[twoside]{article}

\usepackage{csbigs}
\usepackage[utf8]{inputenc}
\usepackage{amsmath}
\usepackage{tikz}
\usepackage{xfrac}
\usetikzlibrary{matrix} 

\makeatletter
\renewcommand{\maketitle}{
\fancypagestyle{plain}{ \fancyhead[L]{Published, CS-BIGS 7(1):14-25}
\fancyhead[C]{}\fancyhead[R]{\url{http://www.csbigs.fr}}\renewcommand
{\headrulewidth}{0pt}}
\thispagestyle{plain}
\bgroup\setlength{\parindent}{0pt}
\begin{flushleft}
  \textbf{\@title}

  \@author
\end{flushleft}\egroup
}
\makeatother

\title{\vspace{10mm}\\
\huge{\textbf{Shrinkage estimation of rate statistics}} 
\vspace{15mm}
} 

\date{}

\author{%
\normalsize \textbf{Einar Holsbø} \\ 
\normalsize \textit{Department of Computer Science, UiT --- The Arctic University of Norway} \\ 
\vspace{1mm}
\normalsize \textbf{Vittorio Perduca}\\ 
\normalsize \textit{Laboratory of Applied Mathematics MAP5, Université Paris Descartes} \\ 
}

\newcommand{\revf}[1]{#1}

\newcommand*\diff{\mathop{}\!\mathrm{d}}

\begin{document}
\sloppy
\maketitle


\renewcommand{\abstractname}{}
\begin{abstract}
This paper presents a simple shrinkage estimator of rates based on Bayesian methods. Our focus is on crime rates as a motivating example. The estimator shrinks each town's observed crime rate toward the country-wide average crime rate according to town size. By realistic simulations we confirm that the proposed estimator outperforms the maximum likelihood estimator in terms of global risk. We also show that it has better coverage properties. 

\vspace{3mm}
\textnormal{Keywords} : Official statistics, crime rates, inference, Bayes, shrinkage, James-Stein estimator, Monte-Carlo simulations. 
\end{abstract}

\vspace{10mm}


\begin{multicols}{2} 

\section{Introduction} 
\subsection{Two counterintuitive random phenomena}
It is a classic result in statistics that the smaller the sample, the more variable the sample mean. The result is due to Abraham de~Moivre and it tells us that the standard deviation of the mean is $\sigma_{\bar{x}} = \sfrac{\sigma}{\sqrt{n}},$ where $n$ is the sample size and $\sigma$ the standard deviation of the random variable of interest. Although the equation is very simple, its practical implications are not intuitive. \textit{People have erroneous intuitions about the laws of chance,} argue Tversky and Kahneman in their famous paper about the law of small numbers \citep{tversky1971belief}. 

\bigskip

Serious consequences can follow from small-sample inference ignoring deMoivre's equation. \citet{wainer2007most} provides a notorious example: in the late 1990s and early 2000s private and public institutions provided massive funding to small schools. This was due to the observation that most of the best schools---according to a variety of performance measures---were small. As it turns out, there is nothing special about small schools except that they are small: their over-representation among the best schools is a consequence of their more variable performance, which is counterbalanced by their over-representation among the worst schools. The observed superiority of small schools was simply a statistical fluke.

\bigskip 

\revf{\citet{galton1886regression} first described another stochastic mechanism that is dangerous to ignore. Galton observed that children of tall (or short) parents usually grow up to be not quite as tall (or short), i.e.\ closer to average height. Today we know this phenomenon as regression to the mean, and we will find it wherever we find variation. Imagine a coach who berates a runner who had an unusually slow lap time and finds that, indeed, the next lap is faster. The coach, who always berates slow runners, has not had the opportunity to realize that the next lap is very likely to be faster no matter what. As long as there is variability in lap time we will some times see unusually slow laps that we can do nothing about and make no inference from. In this case too do our intuitions about the laws of chance fail us. People, including scientists, make the mistake of ignoring regression all the time. Mathematically regression to the mean is as simple as imperfect correlation between instances.}

\subsection{These phenomena in official statistics}
\revf{The small-schools example is egregious because it led to wasteful public spending. The statistics themselves were probably fine, but their interpretation was not careful enough. Such summary statistics are often presented without regard for uncertainty. For instance, every year Statistics Norway (\url{ssb.no}), the central bureau of statistics in Norway, presents crime report counts. The media usually reports these numbers as rates and inform us that some small town that few people know about is the \emph{most criminal} in the country. Often the focus is on violent crimes. Figure \ref{figure:funnel} below shows these rates for 2016. Not knowing de~Moivre's result it might be striking to observe that many of the towns with the highest rates are small towns. Similarly, not knowing regression  it might be striking to observe that, on average, towns with a high rate in one year will have a lower one in any other year, see Figure \ref{figure:regression} below. These are unavoidable stochastic phenomena. Thus there is reason to believe that we should somehow adjust our expectations about these numbers. We will see below that such an adjustment also makes statistical sense.}

\subsection{Shrinkage estimation}
\revf{There is an astonishing decision-theoretic result due to Charles Stein: 
suppose that we wish to estimate $k \geq 3$ parameters $\theta_1, \ldots, \theta_k$ and observe $k$ independent measurements, $x_1\ldots x_k$, such that $x_i \sim N(\theta_i, 1).$ There is an estimator of $\theta_i$ that has uniformly lower risk, in terms of total quadratic loss, than the obvious candidate $x_i$ \citep{stein1956}. In other words, the maximum likelihood estimate is inadmissible. Stein showed this by introducing a lower-risk estimator that biases or \textit{shrinks}, the $x_i$s toward zero. \citet{james1961estimation} introduced an improved shrinkage estimator, which we will see below. \citet{efron1973stein} show a similar result and a similar estimator for shrinking toward the pooled mean. 
There are many successful applications of shrinkage estimation, see for instance the examples from \citet{morris1983parametric}. The common theme is a setting where the statistician wants to estimate many similar variable quantities.}

\subsection{An almost-Bayesian estimator}
\revf{
In this case study we consider the official Norwegian crime report counts. We assume that in a given year the number of crimes reported in town $i$, denoted $k_i$, corresponds to the number of criminal events in this town. We further assume that each inhabitant can at most be reported for one crime a year. Our goal is to estimate the \textit{crime probability} $\theta_i$:  probability that a person will commit a crime in this town. The obvious estimator is the maximum likelihood estimate (MLE) for a binomial proportion $\hat \theta_i = \sfrac{k_i}{n_i},$ where $n_i$ is the population of town $i$.}

\bigskip

\revf{The MLE binomial model rests on an assumption that inhabitants commit crimes independently according to an identical crime probability. There are reasons to believe that this is not the case. The desperately poor might be more prone to stealing than the middle class professional. There is a weaker assumption called \textit{exchangeability} that says that individuals are similar but not identical. More precisely we assume that their \textit{joint} criminal behavior (some number of zeros and ones) does not depend on knowing who the individuals are (the order of the zeros and ones). It is an important theorem in Bayesian inference, due to  De~Finetti, that a sequence of exchangeable variables are independent and identically distributed conditional on an unknown parameter $\theta_i$ that is distributed according to an a priori (or prior) distribution $f(\theta_i)$ \citep{spiegelhalter2004bayesian}. In the binomial sense, $\theta_i$ has the remarkable property that it is the long-run frequency with which crimes occur regardless of the i.i.d.\ assumption; the prior precisely reflects our opinion about this limit. By virtue of De~Finetti's theorem, the exchangeability assumption justifies the introduction of the unknown parameter $\theta_i$ in a binomial model for $k_i$, so long as we take the prior into account.}

\bigskip

\revf{To make an argument with priors is to make a Bayesian argument. 
Shrinkage is implicit in Bayesian inference: observed data gets pulled toward the prior (and indeed the prior is pulled toward the data likelihood). We propose an almost Bayesian shrinkage estimator, $\hat \theta^s_i$, that accounts for the variability due to population size. Our estimator is \emph{almost} Bayesian because we do not treat the prior very formally, as will be clear below.}

\bigskip

\revf{In a Bayesian argument we treat $\theta_i$ as random. The statistician specifies a prior distribution $f(\theta_i)$ for the parameter that reflects her knowledge (and uncertainty) about $\theta_i$. As in the frequentist setting, she then selects a parametric model for the data given the parameters, which allows her to compute the likelihood $f(x|\theta_i)$. Inference about $\theta_i$ consists of computing its posterior distribution by Bayes' theorem:
$$
f(\theta_i|x)=\frac{f(x|\theta_i)f(\theta_i)}{\int f(x|\theta_i)f(\theta_i)\diff\theta_i}.
$$
}

\bigskip

\revf{There are various assessments we could make about the collection of $\theta_i.$ If we assume they are identical we can pool them and use a single prior. If we assume they are independent we specify one prior for each and keep them separate. If we assume they are exchangeable---similar but not identical---it follows from De~Finetti that there is a common prior distribution conditional on which the $\theta_1,\ldots,\theta_m$ are i.i.d.\ \citep{spiegelhalter2004bayesian}.}

\bigskip

\revf{We make this latter judgment and take a beta distribution common to all crime probabilities as prior. Our likelihood for an observed number of crime reports follows a binomial distribution. It is a classic exercise to show that the posterior distribution of $\theta_i$ is then also a beta distribution. The problem remains how to choose the parameters for the prior. On the idea that a given town is probably not that different from all the other towns, we will simply pool the observed crime rates for all towns and fit a beta distribution to this ensemble by the method of moments.}

\bigskip

\revf{Under squared error loss, the posterior mean as point estimate minimizes Bayes risk.} The posterior mean serves as our shrinkage estimate, $\hat\theta^s_i$, for $\theta_i$. We will see that $\hat\theta^s_i$ in effect shrinks the observed crime rate $\hat\theta_i$ toward the country-wide mean $\bar \theta = \sum \frac{1}{m} \hat\theta_i$ by taking into account the size of town $i$.

\bigskip

Bayesian inference allows for intuitive uncertainty intervals.  In contrast to a classical frequentist confidence interval, which can be tricky to interpret, we can say that $\theta_i$ lies within the Bayesian credible interval with a certain probability. This probability is necessarily subjective, as the prior distribution is subjective. We will conduct simulations to compare the coverage properties of our estimator to the classical asymptotic confidence interval.

\subsection{Resources}
This case-study is written with a pedagogical purpose in mind, and can be used by advanced undergraduate and beginning graduate students in statistics as a tutorial around shrinkage estimation and Bayesian methods. We will mention some possible extensions in the conclusion that could be the basis for student projects. Data and code for all our analyses, figures, and simulations are available at \url{https://github.com/3inar/crime\_rates}


\section{Data}

\begin{figure}[H]
  \centering
  \includegraphics[width=0.5\textwidth]{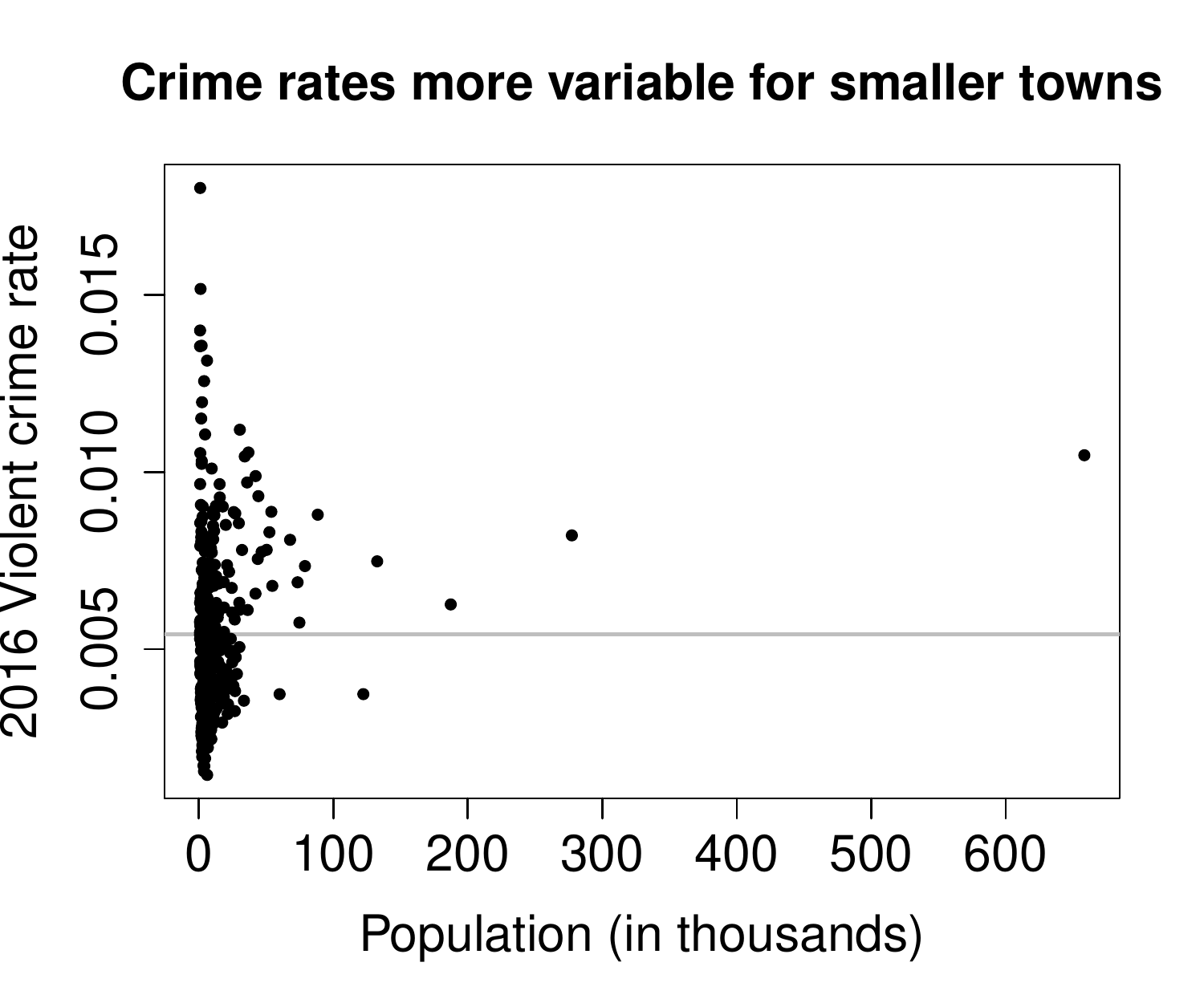}
  \caption{Rates of violent crime vs population in 2016 for all towns in Norway. The grey line shows the country-wide mean.}\label{figure:funnel}
\end{figure}

We will work with the official crime report statistics released by Statistics  Norway (SSB) every year. These data contain the number of crime reports in a given Norwegian town in a given year. The counts are stratified by crime type, e.g.\ violent crimes, traffic violations, etc. We will focus on violent crimes. SSB separately provides yearly population statistics for each town.
Figure \ref{figure:funnel} shows the 2016 crime rates (i.e.\ counts per population) for all towns in Norway against their respective populations. This is some times called a funnel plot for the funnel-like tapering along the horizontal axis: a shape that signals higher variance among the smaller towns.

\begin{figure}[H]
  \centering
  \includegraphics[width=0.5\textwidth]{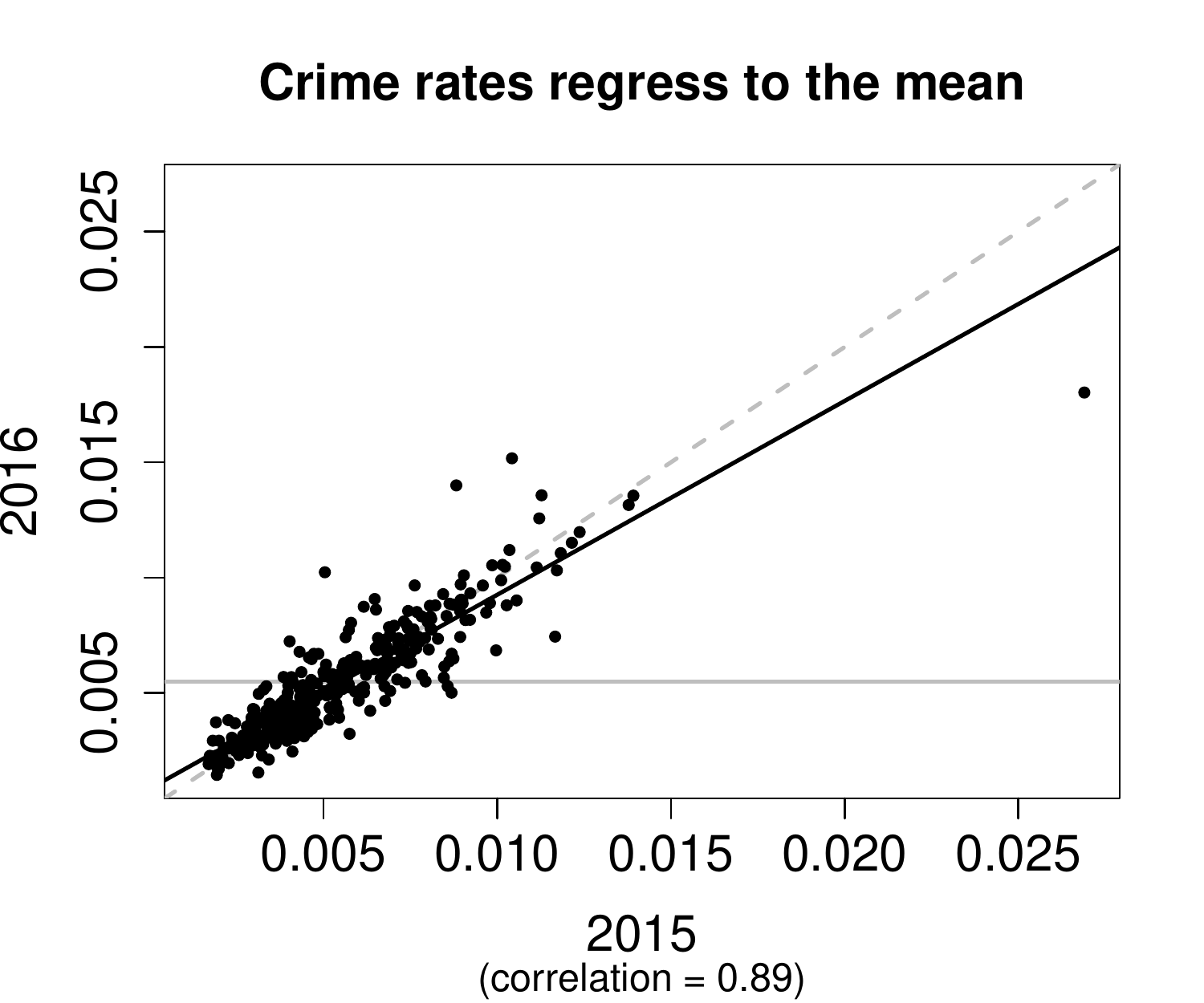}
  \caption{Regression to the mean from year to year. The plot compares 2016 and 2015; the black regression line shows that towns with high crime rates in 2015 tend to have lower crime rates in 2016, and vice versa for low crime rates. The grey dashed line shows what perfect correlation between 2015 and 2016 would look like.}\label{figure:regression} 
\end{figure}

Figure \ref{figure:regression} compares the crime rates in 2015 with those in 2016 and shows that the more (or less) violent towns in 2015 were on average less (or more) violent in 2016. The solid black line regresses 2016 rates on 2015 rates. The dashed grey line is what to expect if there were no regression toward the mean. It has an intercept of zero and a slope of unity. The solid grey line is the overall mean in 2016. The most extreme town in 2015, past .025 on the x-axis, is much closer to the mean in 2016. The solid black regression line shows that this is true for all towns on average. The fact that 2015 and 2016 are consecutive years is immaterial; regression to the mean will be present between any two years.

\begin{figure}[H]
  \centering
  \includegraphics[width=0.5\textwidth]{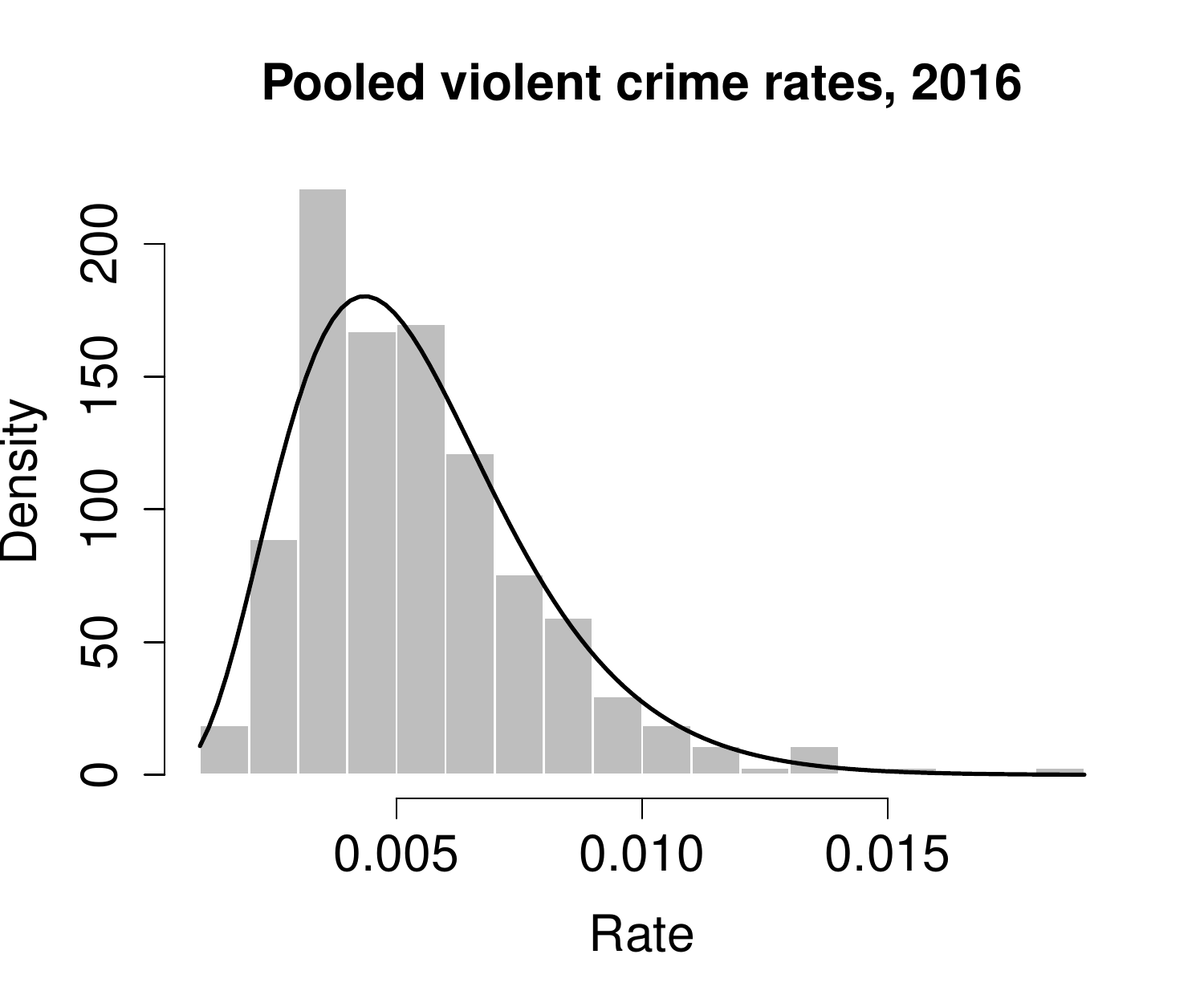}
  \caption{The distribution of violent crime rates in Norway, 2016. The black line describes the method-of-moments fit of a beta distribution to these data.}\label{figure:histogram}
\end{figure}

Figure \ref{figure:histogram} shows the distribution of the pooled violent crime rates for 2016. The solid black line is a beta distribution fit to these data.

\subsection{Simulation study}
\begin{figure}[H]
  \centering
  \includegraphics[width=0.5\textwidth]{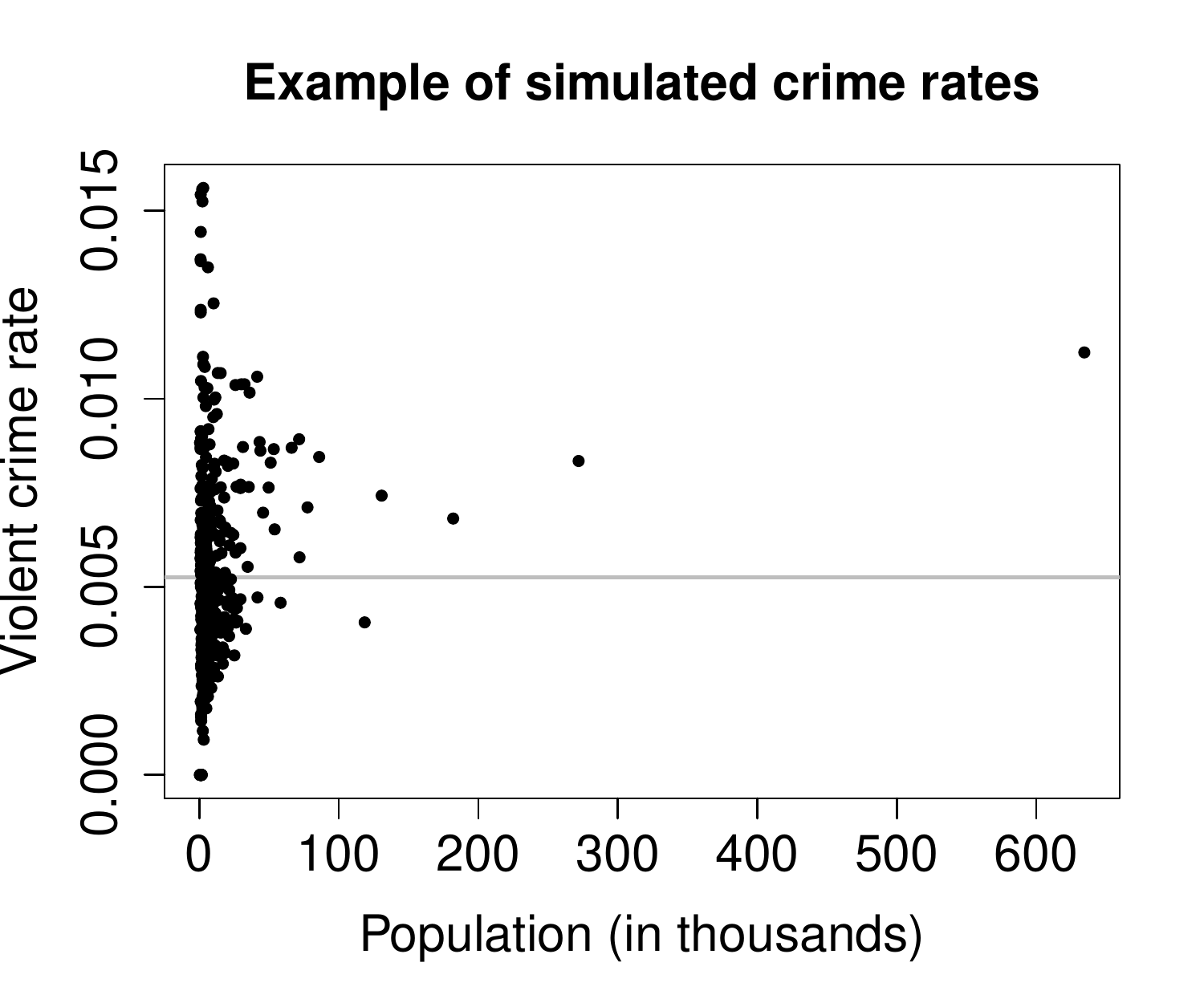}
  \caption{Funnel plot of a set of simulated crime rates}
  \label{figure:funnel_sim}
\end{figure}

\revf{We run a simulation study for validation. If we assume that the crime probability in town $i$ is stationary we can pool the observed crime rates of all years and use their average, $\bar\theta_i$, as a reasonable ``truth.''} This allows us to assess the performance of our estimator against known, realistic crime probabilities, which of course is impossible in the real data. \revf{The simulated crime report count in town $i$ is $k_i \sim \mathrm{Binomial}(\bar\theta_i, n_i),$ where $n_i$ is the 2016 population of town $i$.} Figure \ref{figure:funnel_sim} shows a realization of this procedure. Although not a perfect replica of Figure \ref{figure:funnel}---the real data do not have any rates below .0017---it looks fairly realistic.
 
\section{Methods}
\subsection{Shrinkage estimates}
\begin{figure}[H]
	\centering
	\begin{tikzpicture}

	\matrix[matrix of math nodes, column sep=20pt, row sep=20pt] (mat)
	{
    	    	 & 		    & \alpha,\beta 	&   \\
    	\theta_1 & \theta_2 & \ldots        & \theta_m  \\
    	k_1      & k_2      &  \ldots       & k_m \\
	};
	\draw[->,>=latex] (mat-1-3) -- (mat-2-1);
	\draw[->,>=latex] (mat-1-3) -- (mat-2-2);
	\draw[->,>=latex] (mat-1-3) -- (mat-2-4);
	\draw[->,>=latex] (mat-2-1) -- (mat-3-1);
	\draw[->,>=latex] (mat-2-2) -- (mat-3-2);
	\draw[->,>=latex] (mat-2-4) -- (mat-3-4);
	\end{tikzpicture}
	\caption{\revf{A graph describing our model. Crime counts, $k_i$, are (conditionally) i.i.d.\ binomials whose respective parameters, $\theta_i$, are (conditionally) i.i.d.\ according to a common prior.}}
  	\label{figure:hier_model}
\end{figure}
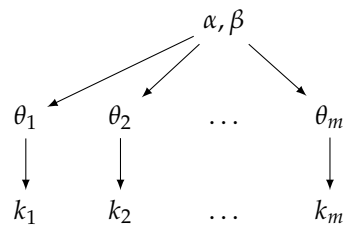

We treat $\theta_i$ as the probability for a person to commit a crime in a given period. We model the total number of crime reports in the $i$-th town, $k_i$, as the number of successful Bernoulli trials among $n_i$, where $n_i$ is the population of this town. As explained in the introduction, this suggests the following simple Bayesian model\revf{, also shown in Figure \ref{figure:hier_model}:} 
\begin{equation*}
\begin{split}
\theta_i\revf{|\alpha,\beta} \sim \mathrm{Beta}(\alpha, \beta), \\
k_i | \theta_i \sim \mathrm{Binomial}(n_i, \theta_i).
\end{split}
\end{equation*}
\revf{As mentioned the assumption of town exchangeability leads to this hierarchical model. This assumption might not be appropriate if we had reasons to think, for instance, that some regions are more prone to crime than others. In this case, region-specific priors might be better.}

The posterior follows from the fact that the beta distribution is conjugate to itself with respect to the binomial likelihood. Generally, conjugacy means that the prior and posterior distributions belong to the same distributional family and usually entails that there is a simple closed-form way of computing the parameters of the posterior. \citet[p.~178]{Wasserman:2010:SCC:1965575} \revf{shows a derivation of the posterior in the beta--binomial model: 
$$
\theta_i | k_i \sim \mathrm{Beta}(\alpha + k_i, \beta + n_i - k_i).
$$
We will look into the relation between the parameters of the posterior to those of the prior in terms of successes and failures in the results section.}

\bigskip

The shrinkage estimate for the crime probability in town $i$ is the posterior mean
$$
\hat\theta^s_i = \frac{\alpha+k_i}{\alpha+\beta+n_i}.
$$
The maximum likelihood estimate for $\theta_i$ is the observed crime rate $\hat \theta_i = \sfrac{k_i}{n_i}$. In order to fix values of $\alpha$ and $\beta$, we pool the MLEs for all towns $\hat\theta_1, \ldots, \hat\theta_m$ and fit a beta distribution to these data by the method of moments. \revf{We show the resulting fit in Figure \ref{figure:histogram}.} Because the expectation and variance of a $\mathrm{Beta}(\alpha, \beta)$ are $\frac{\alpha}{\alpha+\beta}$ and $\frac{\alpha\beta}{(\alpha+\beta)^2(\alpha+\beta+1)}$, respectively, the parameter estimates for the prior are 
\begin{align*}
\beta &= \frac{\alpha(1-\bar{\theta})}{\bar{\theta}}\textrm{, and} \\
\alpha &= \left(\frac{1-\bar{\theta}}{S^2}-\frac{1}{\bar{\theta}}\right)\bar{\theta}^2.
\end{align*}
Here $\bar{\theta}=\frac{\sum_i\hat\theta_i}{m}$ and $S^2=\frac{\sum_i(\hat\theta_i-\bar{\theta})^2}{m-1}$ are the sample mean and variance of the pooled MLEs.

\bigskip

\revf{Instead of estimating $\alpha$ and $\beta$ from the data like this, which ignores any randomness in these parameters, we could have a prior distribution for the parameters themselves. This would yield a typical Bayesian hierarchical model. Note also that in forming the estimate for town $i$, we end up using its information twice: once in eliciting our prior and once in the likelihood. This is convenient because we need only to find one prior rather than one for each town where we exclude the $i$th town from the $i$th prior. This bit of trickery does not make much difference: we have several hundreds of towns and hence removing a single town does not affect the shape of the prior much.}

\bigskip

The estimate $\hat\theta^s_i = \frac{\alpha + k_i}{\alpha+\beta+n_i}$ shrinks the observed, or MLE, crime rate toward the prior mean $\bar\theta$. We can rewrite so that $\hat\theta^s_i = \delta_i\bar\theta + (1-\delta_i)\hat\theta_i$, with
$\delta_i = \frac{\alpha+\beta}{\alpha+\beta+n_i}$. Here $\delta_i$ directly reflects the prior's influence on $\hat \theta^s_i$, and we see that this influence grows as the town size, $n_i$, shrinks.

\revf{
\subsection{James-Stein estimates}
For completeness we demonstrate empirically that the James--Stein estimator is superior to the MLE in terms of risk. If town $i$ has a large enough population, we can consider the normal approximation to the binomial distribution and assume 
$$
\hat\theta_i=\frac{k_i}{n_i}\sim\mathcal{N}\left(\theta_i,\sigma_i^2\right),
$$
where $\sigma_i^2=\frac{\theta_i(1-\theta_i)}{n_i}$ is unknown. If we assume that towns are similar in terms of variance we can consider the pooled variance estimate 
$$
\sigma^2_P=\frac{\sum_{i=1}^m(n_i-1)\hat{\sigma}_i^2}{\sum_{i=1}^m(n_i-1)},
$$
where $\hat{\sigma}^2_i=\frac{\hat{\theta}_i(1-{\hat{\theta}}_i)}{n_i}=\frac{k_i(n_i-k_i)}{n_i^3}$. The James-Stein estimator of crime probability for town $i$ is then 
$$
\hat{\theta}_i^{JS} = \left(1 - \frac{(m-2)\hat\sigma_P^2}{\sum_{i=1}^m \hat\theta_i^2} \right)\hat\theta_i.
$$
This is a shrinkage toward zero. It assumes that crime rates are probably not as high as they appear. This is different from our assumption that crime rates are probably not as far away from the average as they appear. It is simple to modify the above to shrink toward any origin. The Efron-Morris variant \citep{efron1973stein} shrinks toward the average:
$$
\hat{\theta}_i^{JS} = \bar\theta + \left(1 - \frac{(m-2)\hat\sigma_P^2}{\sum_{i=1}^m (\hat\theta_i - \bar \theta)^2} \right)(\hat\theta_i - \bar \theta).
$$
We will use this variant so that the two methods shrink toward the same point.}

\subsection{Uncertainty intervals}
We construct credible intervals from the posterior. A $95\%$ credible interval contains $.95$ of the posterior density, and the simplest way to construct one is to place it between the $.025$ and $.975$ quantiles of the posterior. \revf{For the MLE we use the typical normal approximation (or Wald) confidence interval. There is to our knowledge no straight-forward way to construct confidence intervals for the JS estimator, so we will leave this as an exercise for the reader.}

\subsection{Global risk estimates}
We use the total squared-error loss function,
$$
L(\theta,\hat\theta^s)=\sum_{i=1}^m(\theta_i-\hat{\theta}^s_i)^2,
$$
to measure the global discrepancy between the true rates $\theta=(\theta_i)_{i=1,\ldots,m}$ and estimates $\hat{\theta}^s=(\hat{\theta}_i^s)_{i=1,\ldots,m}$. We do the same for the maximum likelihood and James-Stein estimates $\hat{\theta}=(\hat{\theta}_i)_{i=1,\ldots,m}$ and $\hat{\theta}^{JS}=(\hat{\theta}_i^{JS})_{i=1,\ldots,m}$, respectively. 

\bigskip

We will compare the expected loss, or risk, of the three estimators $R(\cdot)=E[L(\cdot)],$ \revf{confirming the well-known property that shrinkage estimators dominate the MLE}. We obtain Monte Carlo estimates of risk by averaging $L(\cdot)$ across repeated simulations.

\subsection{Coverage properties}
For the credible interval $C^{s} = (a, b),$ we want to assess the coverage probability $\mathbb P(\theta \in C^{s})$ and compare with $\mathbb P(\theta \in C^{W})$ for the classical Wald confidence interval. We will not assess the James--Stein estimator in terms of coverage. Let $I(C_i)$, where $C_i=C_i^{s}$ or $C_i^W$, be the indicator function that is equal to unity if $\theta_i \in C_i,$ and zero otherwise. We obtain MC estimates of coverage probability by averaging the mean internal coverage, $\frac{1}{m} \sum_{i=1}^m I(C_i^{\cdot})$, across repeated simulations. An uncertainty interval should be well-calibrated: if the size of the interval is $95\%$ it should trap the true parameter $.95$ of the time.

\begin{figure}[H]
  \centering
  \includegraphics[width=0.5\textwidth]{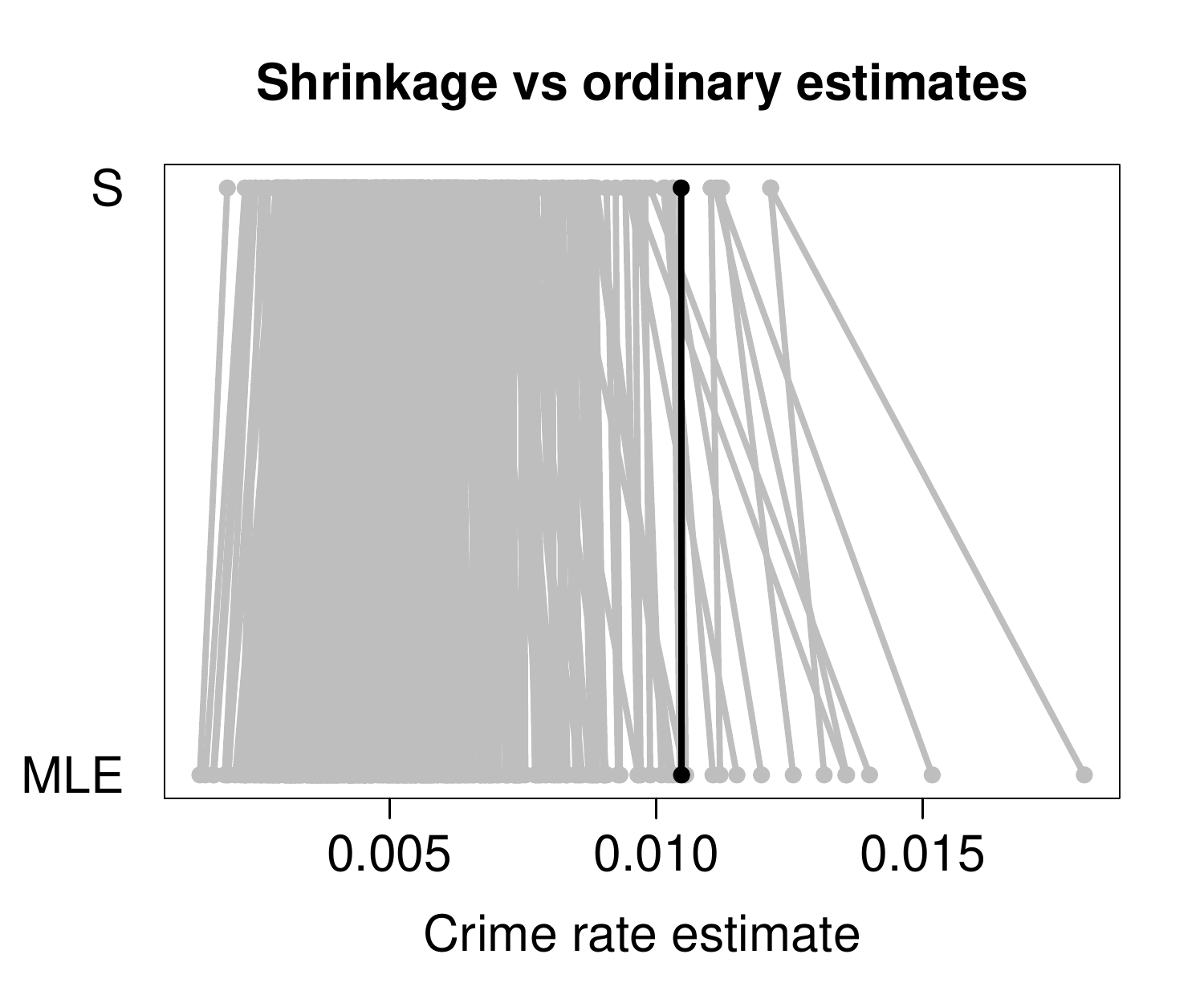}
  \caption{Comparing shrinkage and maximum likelihood estimates. Oslo, in black, is \revf{both close enough to the grand mean and large enough in size that} the estimate does not change.}\label{figure:sticks}
\end{figure}

\section{Results}
\subsection{Official SSB data}
We focus on violent crimes in the year 2016. 
Figure \ref{figure:sticks} shows the effect of shrinking the observed crime rates toward the prior mean. We see that the more extreme estimates shrink toward the center. The city with highest crime rate according to the maximum likelihood estimate is Havsik ($\hat\theta=0.018$), a small town with slightly more than 1000 inhabitants ($n=1054$). After shrinkage, Havsik still ranks first, but the shrinkage estimate is much lower ($\hat\theta^s=0.012$). Similarly the town with the lowest crime rate is Selbu ($\hat\theta=0.0017$), another small town ($n=4132$). Selbu's shrinkage estimate is higher than the MLE by more than 40\%  ($\hat\theta^s=0.0024$). Oslo, shown in black, is a big city ($n=658390$) and the difference between the two estimates is null ($\hat\theta - \hat\theta^s = 7\times 10^{-6}$).

\bigskip
\begin{figure}[H]
  \centering
  \includegraphics[width=0.5\textwidth]{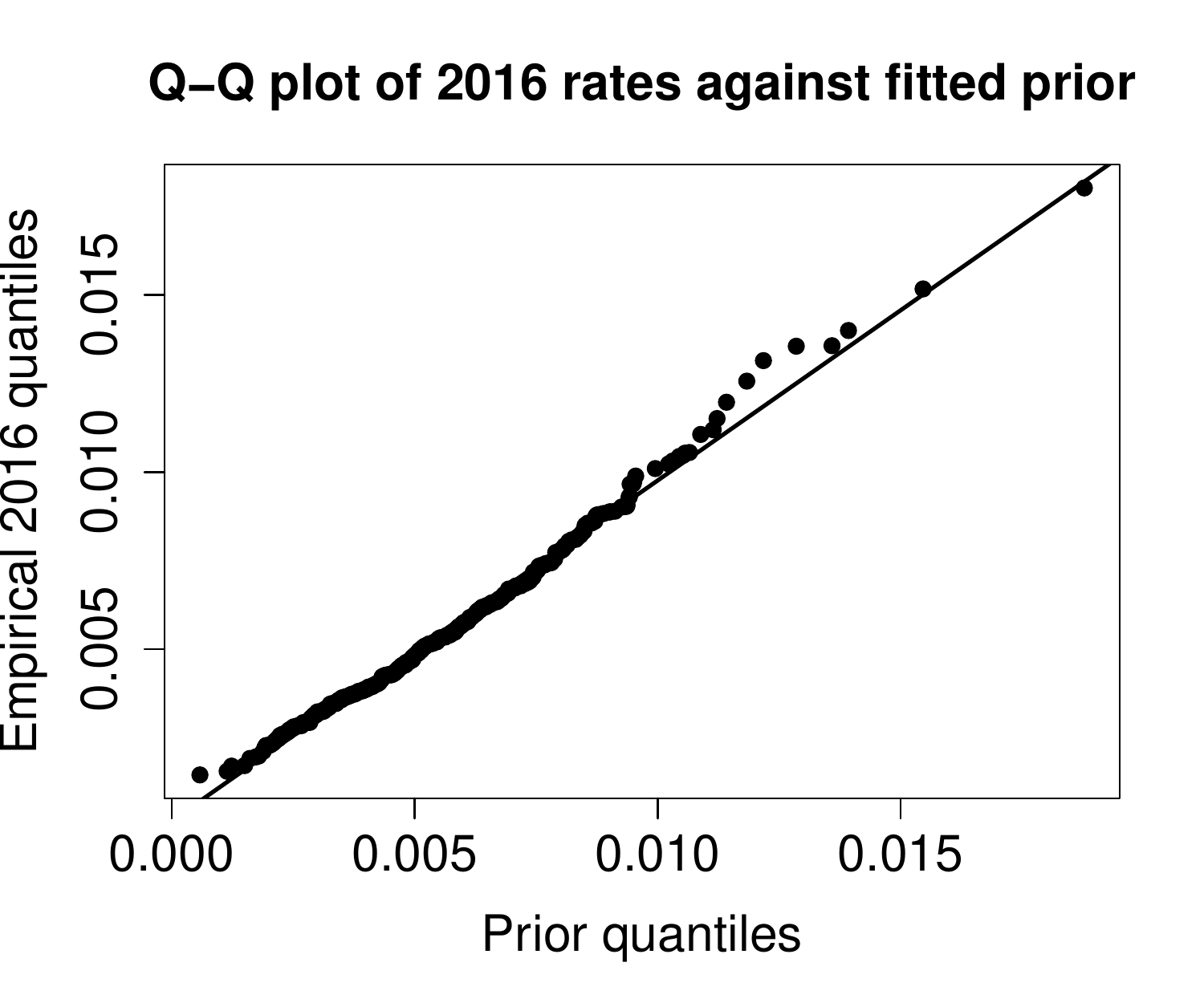}
  \caption{\revf{Quantile--quantile plot of 2016 crime rates against the fitted prior. The solid line describes a perfect fit.}}\label{figure:qqplot}
\end{figure}
\revf{Figure \ref{figure:qqplot} is a quantile--quantile plot of the 2016 violent crime rates against the fitted prior. There is some very slight deviation around the tails, but overall it looks like a nice fit.}

\bigskip

\revf{By shrinking toward the ensemble we add some information---we use the term informally---to the observed rate. We can quantify this by looking at the form of the beta distribution, so far taken for granted in this treatment. Its density function is 
$$f(x; \alpha, \beta) = \frac{x^{\alpha -1}(1-x)^{\beta - 1}}{B(\alpha, \beta)},$$
where the beta function in the denominator is simply the normalizing constant 
$$B(\alpha, \beta) = \int_0^1 t^{\alpha -1}(1-t)^{\beta - 1} \diff t.$$
A natural interpretation is that this is a distribution over the probability of success, i.e.\ crime, in a sequence of Bernoulli trials with $\alpha - 1$ successes and $\beta-1$ failures (cf.\ the binomial distribution). Hence we can interpret the posterior for town $i$ as a distribution over the probability of success in a series of Bernoulli trials with $\alpha^\prime = \alpha + k_i$ successes and $\beta^\prime = \beta + n_i$ failures (ignoring the $-1$ for convenience). In our data we have that $\alpha \approx 5$ and $\beta \approx 917$; it is as though we add the information of 922 extra trials in the binomial sense. In other words we add a priori 922 inhabitants, including five criminals, to each town. 
\begin{figure}[H]
  \centering
  \includegraphics[width=0.5\textwidth]{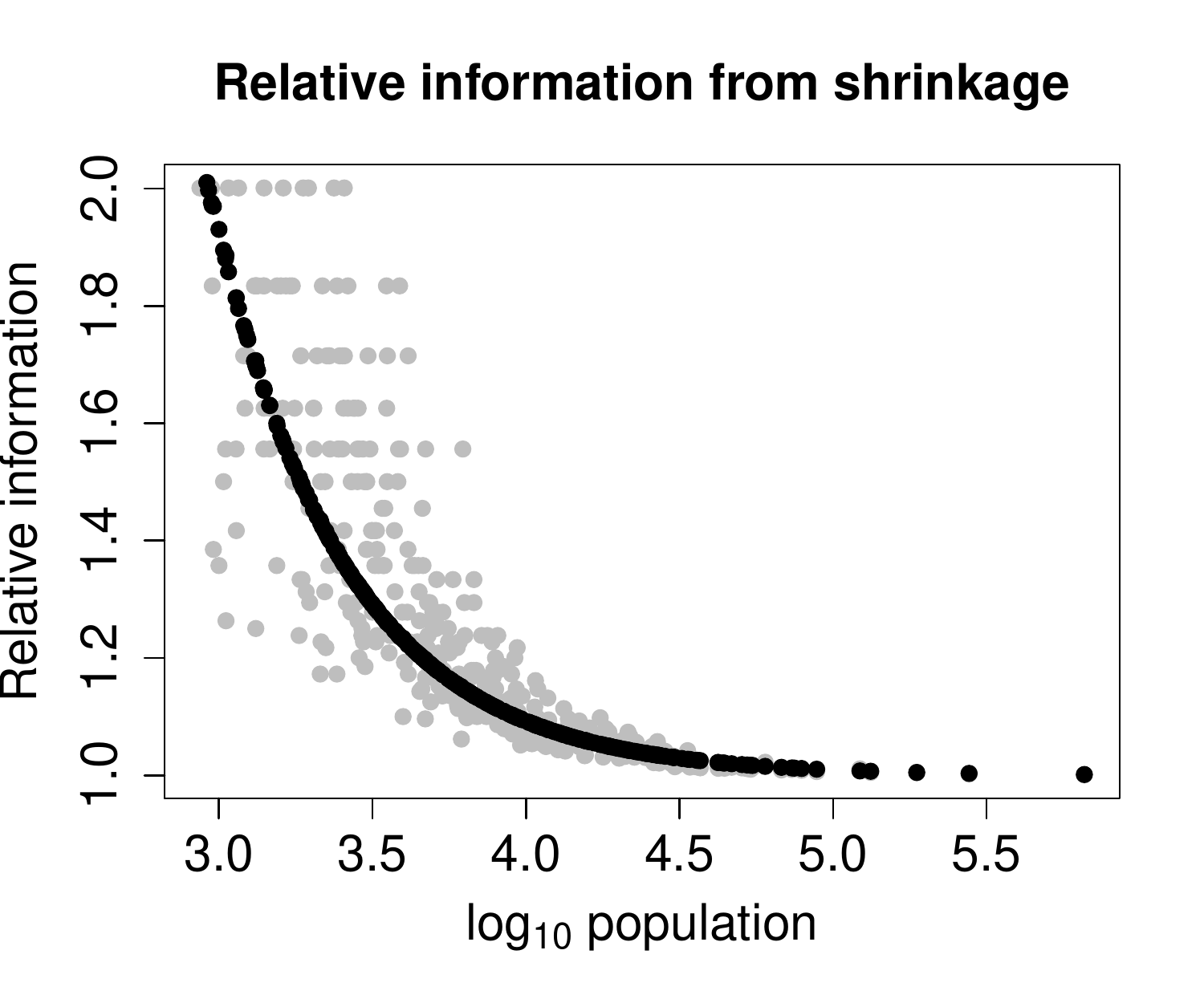}
  \caption{\revf{Relative information in the posterior mean compared to the MLE. The figure shows $(\alpha + k_i)/k_i$ in grey and $(\beta + n_i - k_i)/(n_i - k_i)$ in gray. These represent the added information in terms of number of successes and number of failures added to the MLE to form the shrinkage estimate. For the smallest towns, we practically double the information.}}
  \label{figure:information}
\end{figure}
Figure \ref{figure:information} shows $\alpha^\prime$ and $\beta^\prime$ (gray and black) relative to the number of successes ($k_i$) and failures ($n_i$ - $k_i$) for each town in the 2016 data. For the smaller towns, there is double the information in the shrinkage estimate, while for larger towns there is no practical increase. Naturally the value of this extra information depends on the degree to which the prior is relevant.}

\bigskip

\begin{figure}[H]
  \centering
  \includegraphics[width=0.5\textwidth]{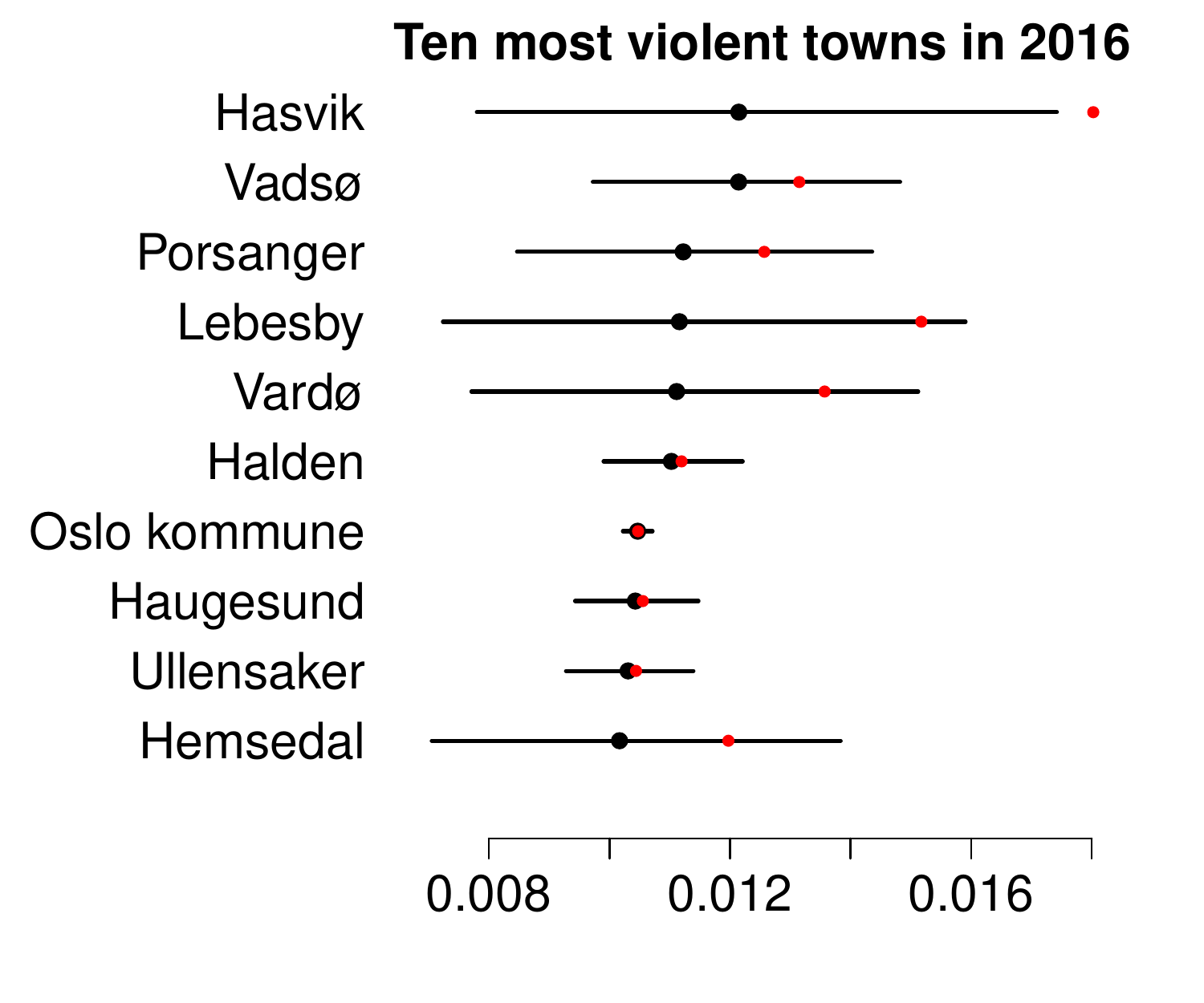}
  \caption{The ten towns with the highest crime rate, ordered by shrinkage estimate. The bars are $95\%$ credible intervals. MLEs shown in red.}\label{figure:topten}
\end{figure}

Figure \ref{figure:topten} shows the ten most violent towns according to shrinkage estimate along with their $95\%$ credible intervals. The official, or MLE, crime rate is shown as a red point. We see some change in ordering. For Hasvik---a small and presumably quiet village in northern Norway---the MLE is so implausible that it is outside the credible interval. For Oslo---the biggest city in Norway---the estimate doesn't change. 

\begin{figure*}[htb]
  \centering
  \includegraphics[width=1\textwidth]{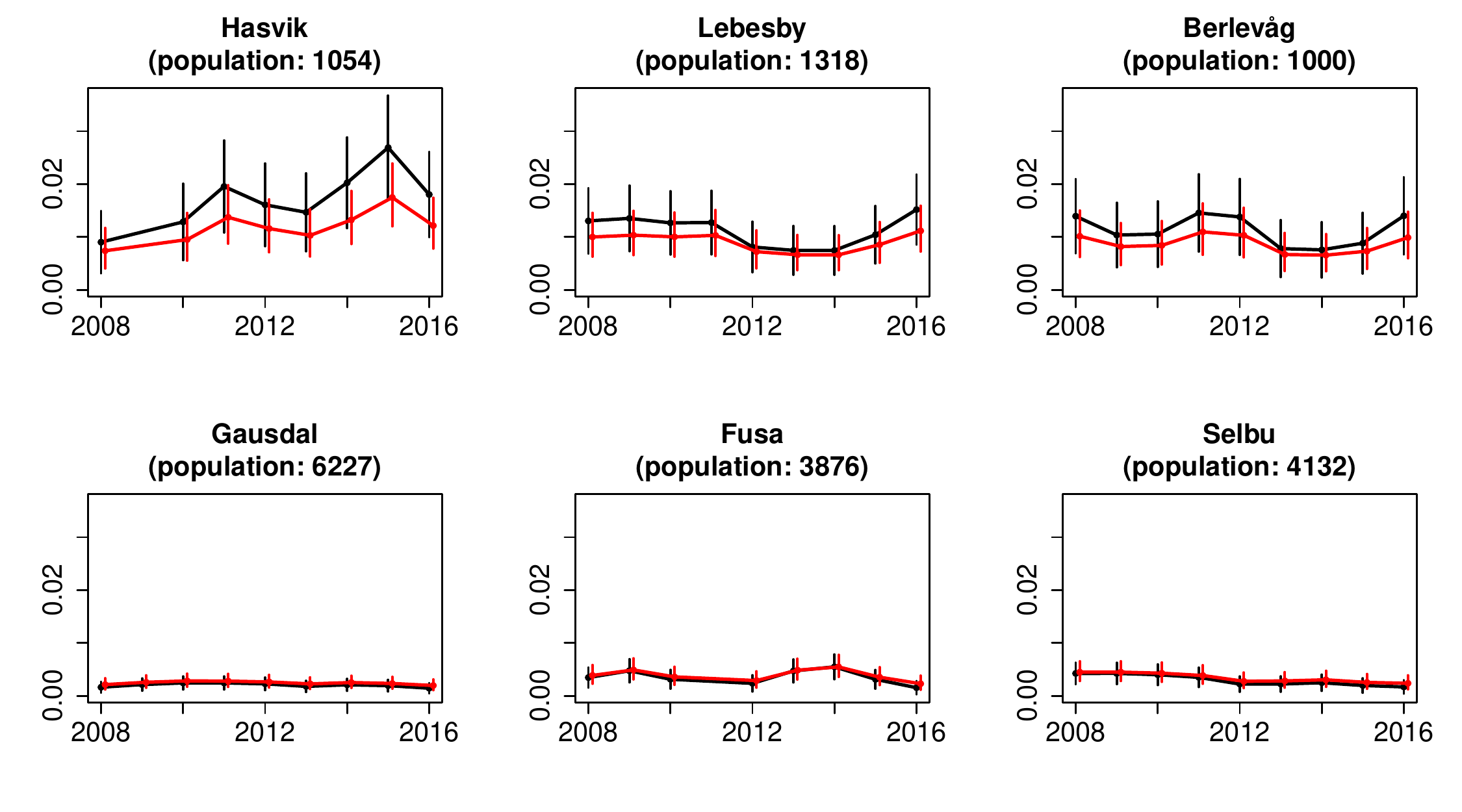}
  \caption{Historical data for the three most violent and the three least violent towns in 2016, ordered by official crime rate (MLE). The official statistics are drawn in black, and shrinkage estimates in red. The vertical bars indicate confidence and credible intervals, respectively.}\label{figure:timeseries}
\end{figure*}

Figure \ref{figure:timeseries}  shows historical data for the three most violent and the three least violent towns in 2016, according to official crime rate. We show shrinkage estimates in red and official statistics in black. The vertical bars are $95\%$ uncertainty intervals. The shrinkage estimate is usually more conservative, at least for the more violent towns, but the trends remain similar for both estimates. The credible intervals are shorter than the classical confidence intervals. We will see that in spite of this their coverage is better under simulation. It is interesting that the three most violent towns are all in Finnmark: Norway's largest and most sparsely populated county.

\bigskip

\subsection{Simulated data}
\begin{figure}[H]
  \centering
  \includegraphics[width=0.5\textwidth]{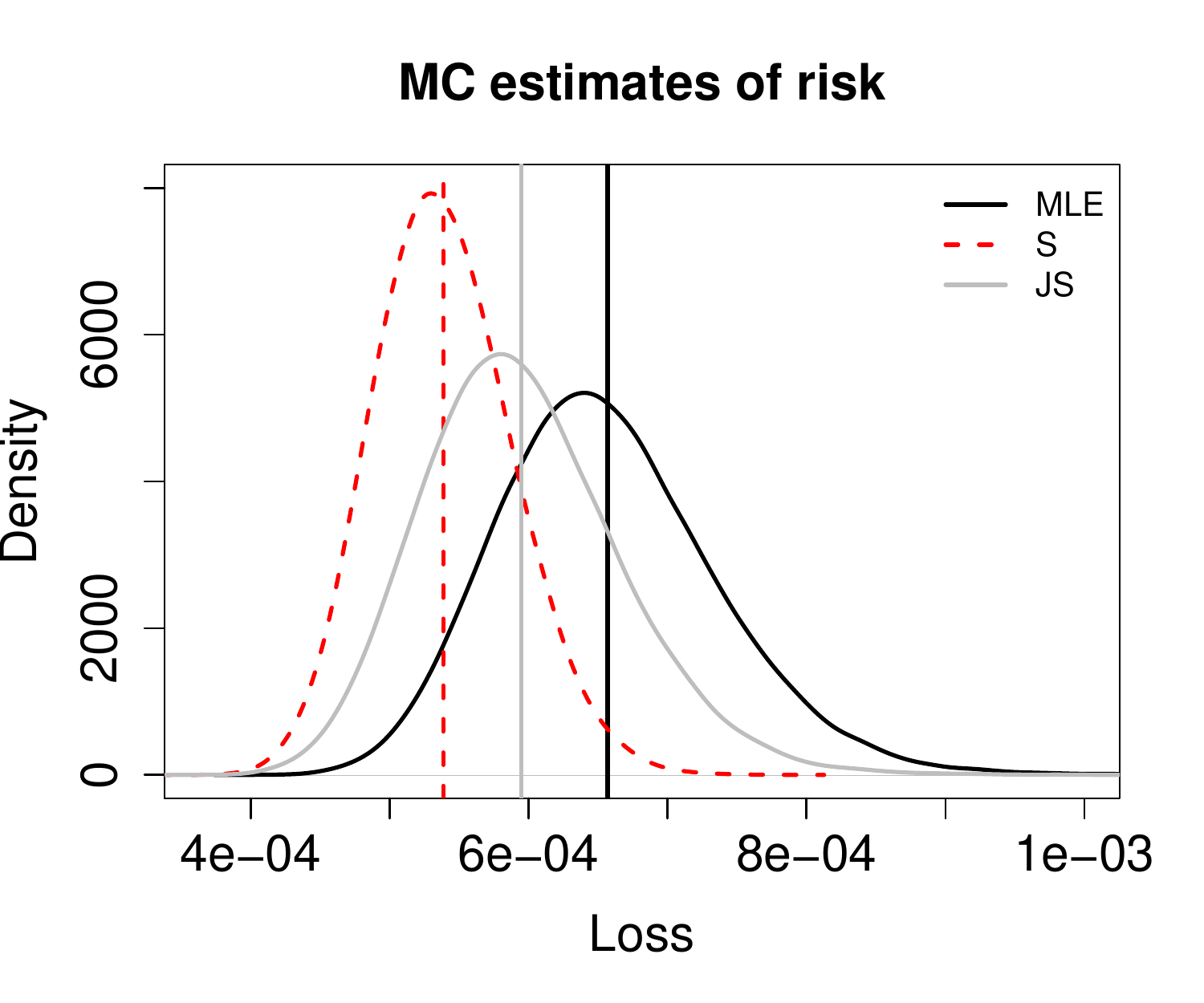}
  \caption{Distributions of $L(\theta,\hat\theta)$ (solid black), $L(\theta,\hat\theta^s)$ (dashed red), and $L(\theta,\hat\theta^{JS})$ (solid grey). Vertical lines estimate the risk.}
  \label{figure:loss}
\end{figure}

To obtain MC estimates of risks we run 100\,000 simulations for each of our two experiments. Figure \ref{figure:loss} shows kernel density estimates of the distributions of global loss. Our shrinkage estimates show lower global risk than maximum likelihood: $\hat{R}(\theta,\hat\theta^s)=0.00054$ versus $\hat{R}(\theta,\hat\theta)=0.00066$. \revf{The James--Stein estimates fall almost exactly between the two with $\hat{R}(\theta,\hat\theta^{JS})=0.00059$. We might have observed better results for JS had we used a variant of JS that allows unequal variances. Note that we fixed $\theta_i$ for this experiment, so we are only assessing the risk function in a single point.}

 \begin{figure}[H]
  \centering
  \includegraphics[width=0.5\textwidth]{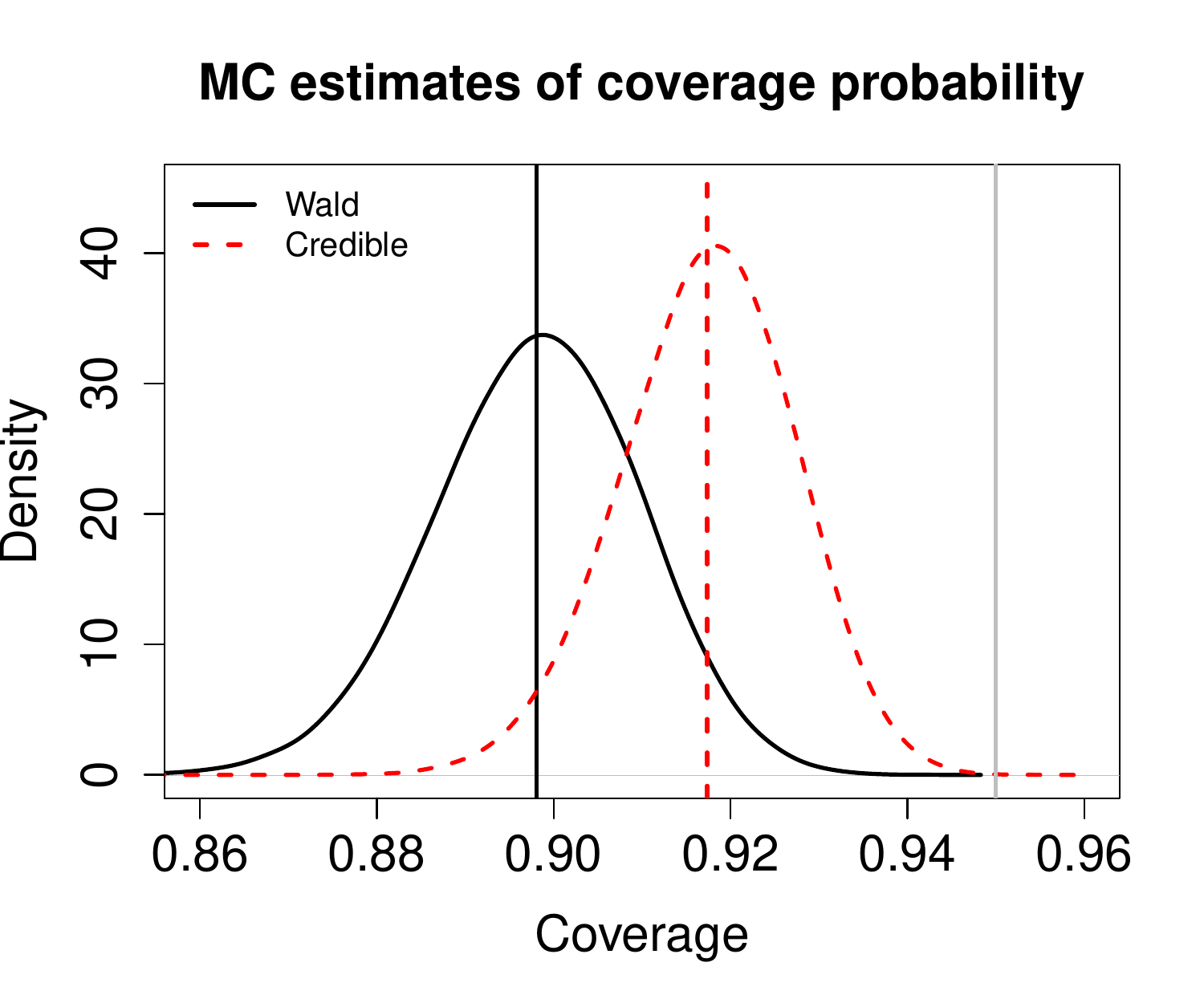}
  \caption{Distributions of the internal coverage $\frac{1}{m} \sum_{i=1}^m I(C_i^{W})$ (solid black) and $\frac{1}{m} \sum_{i=1}^m I(C_i^{s})$(dashed red). Vertical lines estimate coverage probability. The grey line shows the nominal coverage of .95.}
  \label{figure:coverage}
\end{figure}

Figure \ref{figure:coverage} presents estimated coverage probabilities in the same manner as Figure \ref{figure:loss}. The grey line shows the nominal coverage of $.95$. The coverage probability of the credible interval for the shrinkage estimator, $\hat {\mathbb P}(\theta \in C^{s}) = 0.917$, is closer to the nominal value than that of the the standard interval, $\hat {\mathbb P}(\theta \in C^{W}) = 0.898$. There is however still room for improvement.

\section{Conclusion}
This case study shows a simple method for simultaneous estimation of all town-specific crime rates in a country. \revf{The method is Bayesian in spirit, although we take some shortcuts with our prior. It is known that under squared-error loss the posterior mean is the optimal decision w.r.t.\ a given prior. In other words it minimizes Bayes risk, and is called the Bayes estimate. The theory gives us that Bayes estimates are admissible \citep{wald1947}, and thus cannot be dominated. The risk estimates of our simulation agree with this. Our analysis provides an estimate of the crime probability with favorable frequency properties in terms of mean squared error and coverage.}

 
\bigskip

Our simulations show \revf{that the Bayesian credible intervals from this treatment are narrower and have better coverage than the standard Wald confidence interval. Hence we get better information about} the location of $\theta_i$. \citet{brown2001} show extensively that the Wald confidence interval for the binomial proportion behaves erratically for extreme values of $p$, for varying values of $n,$ and for (un)lucky combinations of the two. \revf{Our result is interesting but quite narrow. Generalizing it requires more work.}

\bigskip

Smaller towns are over-represented among the most and least violent towns in the official Norwegian data. Mathematically this has to be the case. Applying shrinkage methods to these data we get more conservative estimates for these variable and often extreme quantities. At the same time it seems that variance is not the only factor that places some of these small towns among the most violent. As Figure \ref{figure:topten} shows, the top and bottom three in 2016 show a certain stability year by year. Hasvik in Finnmark has never ranked especially low since 2008. Small towns in the north are often ranked high for violence. There could be many reasons for this and we leave further analysis to the criminologists.

\bigskip

These simple and useful estimation methods are best understood by practical examples. We encourage readers and students to actively follow this tutorial by playing with the available code and data. We used a single prior for all towns. It would be an interesting extension to use a mixture of beta distributions to account for any heterogeneity due to different latent rate levels. In this case, an EM algorithm could be used to assign each town to a class. Or, since Finnmark seems to be a special case, we might estimate per-county priors. It is also possible to include Bayesian multiple testing procedures to infer a list of cities likely to have true crimes rate above some given threshold. \revf{There is a temporal aspect to these data that we have not looked into. It would be possible to start out with a country-wide prior, but after this let the prior for one year be the posterior from the previous.} Interested readers can find other ideas for further development in \citet{robinson}. \cite{gelman2017teaching} also discuss a similar project to this one in their manual for statistics teachers. 

\bigskip

\revf{In this treatment we have moved from descriptive figures typical of official statistics to model-based inferential statistics, estimating a crime probability rather than reporting a crime count. This allows us to account for variance and perhaps avoid over-interpreting noise, and hence avoid small-schools-type mistakes. We believe that probabilistic thinking can enrich descriptive statistics and aid in their interpretation.}

\section*{Acknowledgements}
\revf{We would like to thank our anonymous reviewer for the very thorough and very useful comments and suggestions.}


\bibliographystyle{apalike}
\bibliography{mybib} 

\end{multicols}

\vspace{25mm}

Correspondence:
\href{mailto:einar@cs.uit.no}{einar@cs.uit.no}

\end{document}